\newcommand{\beq}{\begin{equation}}
\newcommand{\eeq}{\end{equation}}
\newcommand{\bqa}{\begin{eqnarray}}
\newcommand{\eqa}{\end{eqnarray}}
\documentstyle[12pt,pictex,psfig,epsf]{article}
\def\square{\vcenter{\vbox{\hrule height.4pt
          \hbox{\vrule width.4pt height8pt
          \kern8pt\vrule width.4pt}\hrule height.4pt}}}

\begin{document}

\centerline{\Large\bf The Electric Screening Mass in Scalar }
\centerline{\Large\bf Electrodynamics at High Temperature}

\vskip 10mm
\centerline{Jens O. Andersen}
\centerline{\it Institute of Physics}
\centerline{\it University of Oslo}
\centerline{\it P.O. BOX 1048, Blindern}
\centerline{\it N-0316 Oslo, Norway}

\begin{abstract}
{\footnotesize We study the screening of static electric fields
in massless scalar electrodynamics at
high temperature
and zero chemical potential. Effective field theory methods are used
to separate the contributions from the momentum scales 
$T$ and $eT$ to the electric screening mass.
The effects of the distance scale $1/T$ are encoded in the 
parameters of an effective three-dimensional field theory.
The parameters of the effective Lagrangian can be written as a power series
in $e^{2}$. The contribution to physical quantities from the scale
$1/eT$ can be calculated from perturbation theory in the effective
theory and is an expansion in $e$ starting at $e^{3}$. 
The electric screening mass squared 
is computed to order $e^{4}$.}\\ \\ 
PACS number(s): 11.10.Wx, 12.20.DS
\end{abstract}
\section{Introduction}
One interesting phenomenon in hot plasmas is the screening of static
electric fields at long distances. The potential between two static
charges in the plasma
is normally derived in linear response theory~\cite{kapusta},
and reads
\beq
V(R)=Q_{1}Q_{2}\int\frac{d^{3}k}{(2\pi )^{3}}
e^{i{\bf kR}}\frac{1}{k^{2}+\Pi_{00}(0,{\bf k})}.
\eeq
Here, $\Pi_{\mu\nu}(k_{0},{\bf k})$ is the photon polarization tensor.
In the limit $R\rightarrow \infty$, the potential is dominated by the pole in
photon propagator. At leading order this pole is given by the infrared limit
of $\Pi_{00}(0,{\bf k})$, and the potential is thus a modified Coulomb
potential with an inverse screening length 
or electric screening mass $\Pi_{00}(0,{\bf k}\rightarrow 0)$.
This has lead
one to {\it define} the electric screening mass as the infrared
limit of the polarization tensor~\cite{kapusta}:
\beq
m_{s}^{2}=\Pi_{00}(0,{\bf k}\rightarrow 0).
\eeq
This definition cannot be the correct one, since, beyond leading order in
the coupling, it 
is gauge-fixing dependent in non-Abelian theories [2,3]. Although 
$\Pi_{\mu\nu}(k_{0},{\bf k})$ is a manifestly gauge-fixing independent
quantity in Abelian theories, the infrared limit is not renormalization 
group invariant, and is so a useless definition even here. 

The electric screening mass is correctly defined as the 
the position of the pole of the propagator at spacelike 
momentum [2,3]:
\beq
\label{propdef1}
k^{2}+\Pi_{00}(0,{\bf k})=0,\hspace{1cm}k^{2}=-m_{s}^{2}.
\eeq
Here, $k=|{\bf k}|$.
This definition is gauge fixing independent order by order in perturbation
theory, which can be proved on an algebraic 
level~\cite{kobes}\footnote{The pole position
is also independent
of field redefinitions. Since the relation between the fields in the 
underlying theory and the effective theory can be viewed as a field 
redefinition, and since the screening mass is a long-distance quantity, 
one can use the effective theory to compute it. See below.}. 
We also note that the two definitions normally coincide at leading order
in the coupling constant.
The above definition can be extended to other theories, 
e.g. $\phi^{4}$-theory.
The polarization tensor is then replaced by the self-energy function 
for the
scalar field, and the screening mass then reflects the screening of
static scalar fields in the plasma.

However, it turns out that one cannot calculate perturbatively the 
screening
mass beyond leading order in non-Abelian gauge theories using 
Eq.~(\ref{propdef1}) [2,3]. The problem is a linear mass-shell singularity.
This signals the breakdown of perturbation theory, and calls for a 
gauge-fixing independent
and nonperturbative definition of the electric screening mass~\cite{polyakov}.

In Abelian gauge theories the above definition is equivalent to defining the 
Debye mass as
the correlation length of equal-time electric field correlation function
\cite{polyakov}
\beq
\label{corr}
\langle {\bf E}({\bf x})\cdot{\bf E}(0)\rangle\sim
e^{-m_{s}x}/x^{3},
\eeq
where $x=|{\bf x}|$. 
Unfortunately, the
definition Eq.~(\ref{corr}) is a poor one in non-Abelian theories,
since ${\bf E}$ is no longer a gauge invariant quantity.
The above considerations have
lead Arnold and Yaffe \cite{polyakov} to define the electric screening
in terms of Polyakov loops. We shall not pursue this any further, but stick
to the definition based on the pole of the propagator.

As is now well-known, higher order perturbative calculations at finite
temperature requires resummation in order 
to calculate consistently in powers of the coupling constant~\cite{pis}.
The usual connection between the number
of loops in the loop expansion and the powers of coupling constant is
lost in quantum field theories at high temperature for
processes where every external momentum is soft~\cite{pis}. Effects of
leading order in $g$ arise from every order in the loop expansion, and a
resummation of the usual series of loop diagrams is required.
In $\phi^{4}$-theory, resummation simply amounts to replacing the bare
scalar propagator, $1/P^{2}$, by an effective one, $1/(P^{2}+m_{s}^{2})$,
where $m_{s}$ is a thermal mass that goes like $gT$.
This remark also applies to Abelian gauge theories, where the one-loop
self-energy diagram generates an electric screening mass for the timelike
component of the gauge field. 
In non-Abelian gauge theories the situation is somewhat more complicated
in the sense that it is also necessary to use effective vertices with a 
non-trivial momentum dependence, instead of the bare ones.

For the calculations of static quantities, such as free energies and 
screening masses, these are most conveniently carried out in the
imaginary formalism, without analytic continuation to real time~\cite{parw1}.
In this case there exists a simplified resummation scheme due to Arnold and
Espinosa~\cite{arnold}. The point is that in the Euclidean formalism, the
momentum can be soft only for $n=0$. Hence, it is only for the static
modes that one uses effective propagators instead of the bare ones.
Calculations based on this splitting are simpler to carry out than using
the full resummation program of Braaten and Pisarski, and this 
approach has been used
to the study phase transitions in gauge theories~\cite{arnold}, the 
calculation of the free
energy in e.g. $\phi^{4}$-theory and QCD [9,10]
as well 
as in the computation of the electric screening mass in QED and 
SQED~\cite{parw1}.

In the study of static phenomena, such as the screening of static electric
fields, there exists a convenient alternative to resummed 
perturbation theory,
which is based on dimensional reduction and effective field theories.
The idea is that the imaginary time formalism of finite 
temperature field theory involves free propagators in the
form $[{\bf p^{2}}+p_{0}^{2}+m^{2}]^{-1}$ where $p_{0}=2n\pi T$
for bosons and $(2n+1)\pi T$ for fermions
and a summation over
$n$. Thus for distances large compared to $1/T$, one expects that the 
nonstatic
modes decouple, and one is left with an effective three-dimensional field
theory of the zero-frequency modes
of the original fields. One strategy has therefore
been to integrate out the nonstatic modes in some appproxmation. Dimensional
reduction
was first studied in detail in the papers by Ginsparg~\cite{gins} and
Landsman~\cite{lands}, and has later been applied by a number of authors
(\cite{laine} and Refs. therein). 

Instead of literally integrating out the heavy modes, one 
may use an
effective field theory approach to dimensional reduction.
This method has recently been developed 
in an important paper by Braaten and Nieto \cite{braaten}.
One writes down the most 
general 
three-dimensional Lagrangian consistent with the symmetries of 
the system. 
The coefficients
are determined by demanding that the static
correlators in the 
full
theory are reproduced by the ones in the effective theory 
at distances $R\gg 1/T$ to some desired
accuracy.
Braaten and Nieto
applied this method
to $\phi^{4}$ theory for calculating the screening mass
squared to order $g^{5}\ln g$
and the free energy to order $g^{6}\ln g$. Later they
performed a calculation of the free energy in
non-Abelian gauge theories 
\cite{braaten3} to order $g^{5}$, 
confirming the calculations of Zhai and Kastening~\cite{kast}.
These methods have also been applied by the present author 
to hot QED~\cite{jens}.
The calculations
reproduce the results obtained by resummed perturbation
theory for the free energy [10,16] and the screening mass
squared~\cite{parw1}, both to order $e^{5}$.

In the present work we apply the methods of Braaten and Nieto,
and the plan of the article is as follows. In section two we 
discuss SQED at high temperature and the symmetries
of the effective Lagrangian.
In section three, the short distance coefficients
in the effective theory are determined, and in section four we compute
the electric screening mass squared
to order $e^{4}$. Our calculations confirm
the result of Kalahsnikov and Klimov~\cite{klimov}, 
who were the first to compute the screening mass to order $e^{3}$, and
Blaizot {\it et al.}~\cite{parw1}, who have derived it to order
$e^{4}$ by resummation methods.
In section five, we draw some conclusions, 
and the appendices provide the reader
the necessary details as well as our conventions and notation.
In the Feynman graphs a wavy line denotes the photon, a solid line
the complex scalar, and a dashed line the real scalar in the effective
theory.
\section{Scalar Electrodynamics at High Temperature}
The Euclidean Lagrangian of SQED is
\beq
{\cal L}_{\mbox{\scriptsize SQED}}
=\frac{1}{4}F_{\mu\nu}F_{\mu\nu}+({\cal D_{\mu}}\Phi )
^{\dagger}({\cal D_{\mu}}\Phi )
+{\cal L}_{\mbox{\footnotesize gf}}
+{\cal L}_{\mbox{\footnotesize gh}},
\eeq
where we have ignored the scalar self-coupling, since we will focus 
on the electromagnetic interactions.
The covariant derivative is ${\cal D}_{\mu}=\partial_{\mu}+ieA_{\mu}$ and
the gauge fixing term is chosen to be
${\cal L}_{\mbox{\footnotesize gf}}=
\frac{1}{2}(\partial_{\mu}A_{\mu})^{2}$,
which is the Feynman gauge. The ghost term is then 
${\cal L}_{\mbox{\footnotesize gh}}=
(\partial_{\mu}\overline{\eta})(\partial_{\mu}\eta)$
and is thus decoupled from the rest of the Lagrangian.
 
Physical quantities in SQED at high temperature receive contributions 
from the
momentum scale $T$ which is a typical momentum of 
a particle in the plasma and
from the momentum scale $eT$, which is the scale of electric screening.
In non-Abelian theories there is also a contribution from the
momentum scale $g^{2}T$, which is the order of the inverse confinement
radius of QCD in three dimensions.

The two scales in scalar electrodynamics at high temperature
suggests that one should integrate out the scale
$1/T$ and construct an effective three-dimensional Lagrangian, which is
equivalent to SQED at momentum scales up to $eT$.
We call this effective field theory
electrostatic 
scalar quantum electrodynamics (ESQED), in analogy with the 
the definitions introduced by Braaten and Nieto 
in the case of QCD \cite{braaten3}. 
This effective theory consists of 
a real scalar field coupled to SQED in three dimensions. The fields
can be identified, up to normalizations, with the $n=0$ mode of the
fields in the original theory. In particular, the real scalar field is
identified with the zero-frequency mode of the
time-like component of the gauge field. 

Modern developments in renormalization theory implies that static
correlators in SQED at long distances $R\gg 1/T$ can be reproduced
by ESQED to any desired accuracy by adding sufficiently many operators
to ${\cal L}_{\mbox{\scriptsize ESQED}}$ and tuning the coefficients 
as functions of $e$, $T$ and the renormalization scale
$\Lambda$. The dependence of $\Lambda$ in the parameters
is canceled by the $\Lambda$-dependence of loop integrals in ESQED.

Now, ${\cal L}_{\mbox{\scriptsize ESQED}}$ must be a gauge invariant
function of the spatial fields $A_{i}$,
up to the usual gauge fixing terms. This symmetry
follows from the corresponding symmetry in the 
four dimensional theory
and the Ward-Takahashi identity in the high temperature 
limit~\cite{lands}.
The breakdown of Lorentz invariance at finite temperature allows the
the temporal component of the gauge field to acquire a thermal mass.
Moreover, there is a three dimensional
rotational symmetry and a discrete symmetry
$A_{0}\rightarrow -A_{0}$. 
The effective Lagrangian then has the general form
\bqa \nonumber
{\cal L}_{\mbox{\scriptsize ESQED}}&=&
\frac{1}{4}F_{ij}F_{ij}+({\cal D}_{i}\phi )^{\dagger}
({\cal D}_{i}\phi )+M^{2}(\Lambda )
\phi^{\dagger}\phi
+\frac{1}{2}(\partial_{i}A_{0})
(\partial_{i}A_{0})+\\
&&
\frac{1}{2}m^{2}(\Lambda )A_{0}^{2}
+e^{2}_{E}(\Lambda)A_{0}^{2}
\phi^{\dagger}\phi
+{\cal L}_{\mbox{\footnotesize gf}}
+{\cal L}_{\mbox{\footnotesize gh}}
+\delta{\cal L}.
\eqa
Here $\delta {\cal L}$ represents all other terms consistent
with the symmetries.
Examples of such terms
are $\lambda (\Lambda )A_{0}^{4}$ and $\tilde{\lambda}
(\Lambda )(\phi^{\dagger}\phi )^{2}$, which are 
superrenormalizable and $h(\Lambda )(F_{ij}F_{ij})^{2}$, which is 
non-renormalizable.
\section{Parameters in ESQED}
In the next subsections,  we shall determine the short distance
coefficients $e^{2}_{E}(\Lambda)$, $M^{2}(\Lambda)$, and $m^{2}(\Lambda)$ 
We shall do so by calculating various quantities in the two theories
using
strict 
perturbation theory \cite{braaten} and demanding that they match.
Strict perturbation theory is simply ordinary perturbation theory in powers
of $e^{2}$, neglecting resummation. In the full theory we then split the 
Lagrangian into a free part and an interaction part accordingly:
\bqa \nonumber
({\cal L}_{\mbox{\scriptsize SQED}})_{0}&=&\frac{1}{4}
F_{\mu\nu}F_{\mu\nu}+(\partial_{\mu}
\Phi )^{\dagger}(\partial_{\mu}\Phi )
+{\cal L}_{\mbox{\footnotesize gf}}
+{\cal L}_{\mbox{\footnotesize gh}},
\\ \nonumber
({\cal L}_{\mbox{\scriptsize SQED}})
_{\mbox{\footnotesize int}}&=&
e^{2}\Phi^{\dagger}\Phi A_{\mu}^{2}
-ieA_{\mu}
(\Phi^{\dagger}\partial_{\mu}\Phi-\Phi\partial_{\mu}
\Phi^{\dagger}).
\eqa
In the effective theory the mass parameters as well as $e^{2}_{E}(\Lambda)$
are of order $e^{2}$, 
while all other parameters are of order $e^{4}$ or
higher. Thus the masses as well as higher order operators
are treated as perturbations. We then write 
${\cal L}_{\mbox{\scriptsize ESQED}}=
({\cal L}_{\mbox{\scriptsize ESQED}})_{0}
+({\cal L}_{\mbox{\scriptsize ESQED}})
_{\mbox{\footnotesize int}}$
and strict perturbation theory
corresponds to the following partition of the 
effective Lagrangian 
\bqa \nonumber
({\cal L}_{\mbox{\scriptsize ESQED}})_{0}&=&\frac{1}{4}
F_{ij}F_{ij}+(\partial_{i}
\phi )^{\dagger}(\partial_{i}\phi )
+\frac{1}{2}(\partial_{i}A_{0})^{2}
+{\cal L}_{\mbox{\footnotesize gf}}
+{\cal L}_{\mbox{\footnotesize gh}},
\\ \nonumber
({\cal L}_{\mbox{\scriptsize ESQED}})
_{\mbox{\footnotesize int}}&=&\frac{1}{2}m^{2}(\Lambda )A_{0}^{2}+
M^{2}(\Lambda )\phi^{\dagger}
\phi+
e^{2}_{E}(\Lambda)\phi^{\dagger}\phi (A_{i}^{2}+A_{0}^{2})\\
&&-ie_{E}(\Lambda)A_{i}
(\phi^{\dagger}\partial_{i}\phi-\phi\partial_{i}
\phi^{\dagger})+\delta{\cal L}.
\eqa
The expansion in $e^{2}$ in the full four-dimensional theory as well as in the
effective field theory in three dimensions becomes increasingly infrared
divergent as one goes to higher orders in the loop expansion.
However, if one determines the parameters in ESQED so that it is 
equivalent to SQED at distances $R\gg T$, then the infrared dievergences
will also be the same. This implies that one can use a strict perturbation
expansion in $e^{2}$ to determine the parameters of ESQED, although
the infrared problems are not treated correctly~\cite{braaten}.  
\subsection{The Coupling Constant}
For the calculation of the electric screening
mass squared to order $e^{4}$, we need the gauge coupling 
$e_{E}$
only to leading order in $e$. By using the 
relation between the gauge
fields in the two theories
\beq
A_{i}^{3}=\sqrt{T}A_{i},
\eeq
and comparing ${\cal L}_{\mbox{\scriptsize ESQED}}$ 
with $\int_{0}^{\beta}{\cal L}_{\mbox{\scriptsize SQED}}$, we find 
\beq
e_{\footnotesize E}^{2}(\Lambda)=e^{2}T.
\eeq
At this order there is no dependence on the renormalization 
scale $\Lambda$.
\subsection{The Mass Parameters}
In this subsection we calculate the parameters $M^{2}(\Lambda)$ and 
$m^{2}(\Lambda)$ at leading and next-to-leading order 
in $e^{2}$, respectively. 
The physical interpretation of a mass parameter 
is that it is the 
contribution to the physical screening mass from 
momenta of order $T$.
The simplest way of determining the
mass parameters is to match the screening masses in SQED and
in ESQED. Denoting the self-energy
for the field $\Phi$ by $\Sigma (k_{0},{\bf k})$, 
the {\it scalar} screening mass is the solution to the 
equation\footnote{We remind the reader that this screening mass has nothing
to with the screening of electric fields.
As noted in the introduction, this is a quantity which gives information
about the screening of static scalar fields due to rearrangements in the 
plasma.}
\beq
\label{propdef}
k^{2}+\Sigma(0,{\bf k})=0,\hspace{1cm}k^{2}=-m^{2}_{s}.
\end{equation}
The matching requirement implies that 
\beq
\label{sscr}
k^{2}+M^{2}(\Lambda )+\Sigma_{E}(k,\Lambda )=0,
\hspace{1cm}k^{2}=-m^{2}_{s},
\eeq
where $\Sigma_{E}\,(k,\Lambda )$ 
is the self-energy of the field $\phi$ in the effective
theory. 
The self-consistent solution to Eq.~(\ref{propdef}) is to leading order
in the coupling constant
$m_{s}^{2}\approx \tilde{\Sigma}_{1}(0)$, where $\tilde{\Sigma}_{n}(k^{2})
\equiv\Sigma_{n}(0,{\bf k})$ denotes the
nth order contribution to the self-energy in the loop expansion and
the symbol $\approx$ is a reminder that this unphysical screening mass
is obtained in strict perturbation theory.
The relevant diagrams are depicted in Fig.~\ref{1skalar}
and the one-loop self-energy at zero external momentum is given by 
\beq
\tilde{\Sigma}
_{1}(0)=(d-1)e^{2}\hbox{$\sum$}\!\!\!\!\!\!\int_{P}\frac{1}{P^{2}}.
\eeq
Here, $d=4-2\epsilon$ and the sum-integral is defined in appendix A.
This immediately gives
\beq
\tilde{\Sigma}_{1}(0)=\frac{e^{2}T^{2}}{4}.
\eeq
In the effective theory the diagrams which contribute to the self-energy
is shown in Fig.~\ref{eskalar}. Here the blob indicates a mass insertion.
The self-energy function $\Sigma_{E}(k,\Lambda)$ vanishes in strict 
perturbation theory,
since all the propagators are massless. Hence the matching requirement
gives $m_{s}^{2}\approx M^{2}(\Lambda)$ and the mass parameter is
\beq
M^{2}(\Lambda)=\frac{e^{2}T^{2}}{4}.
\eeq
At this order $M^{2}(\Lambda)$ is independent of the scale $\Lambda$.

Let us now turn to the mass parameter $m^{2}(\Lambda)$. The 
screening mass is again defined as the pole of the propagator at spacelike
momentum
\beq
\label{prop}
k^{2}+\Pi_{00}(0,{\bf k})=0,\hspace{1cm}k^{2}=-m^{2}_{s}.
\end{equation}
The self-energy function is given by a series expansion in $e^{2}$ and
can also be expanded in a Taylor series around $k^{2}=0$. 
The self-consistent
solution to Eq.~(\ref{prop}) at next-to-leading order in the coupling
constant is then
\beq
\label{scrm}
m_{s}^{2}\approx\Big[\Pi_{1}(0)+\Pi_{2}(0)\Big]
\Big [1-\Pi_{1}^{\prime}(0)\Big].
\eeq 
Here we have defined
$\Pi(k^{2})\equiv\Pi_{00}(0,{\bf k})$ and $\Pi_{n}(k^{2})$ denotes the 
nth order contribution
to $\Pi (k^{2})$ in the loop expansion. The one-loop self-energy 
is shown in Fig~\ref{selfa}.
and equals
\bqa 
\Pi_{1}(k^{2})&=&2e^{2}\hbox{$\sum$}\!\!\!\!\!\!\int_{P}\frac{1}{P^{2}}
-4e^{2}\hbox{$\sum$}\!\!\!\!\!\!\int_{P}\frac{p^{2}_{0}}{P^{2}(P+K)^{2}} 
\label{divergent}.
\eqa
Expanding in powers of the external momentum and integrating by parts
in $d-1$ dimensions yields
\bqa
\Pi_{1}(k^{2})&=&2e^{2}\hbox{$\sum$}\!\!\!\!\!\!\int_{P}\frac{1}{P^{2}}
-4e^{2}\hbox{$\sum$}\!\!\!\!\!\!\int_{P}\frac{p_{0}^{2}}{P^{4}}
+\frac{4}{3}e^{2}k^{2}\hbox{$\sum$}\!\!\!\!\!\!\int_{P}\frac{p_{0}^{2}}{P^{6}}
+O(k^{4}/T^{2}). 
\eqa
The last sum-integral is ultraviolet divergent and this 
divergence may
be removed by the wave function renormalization counterterm:
\beq
Z_{\mbox{\scriptsize A}}=1-\frac{e^{2}}{3(4\pi)^{2}\epsilon}.
\eeq
One then obtains
\bqa
\label{1loop}\nonumber
\Pi_{1}(k^{2})&=&\frac{e^{2}T^{2}}{3}+\frac{k^{2}}{3(4\pi )^{2}}
(2+2\gamma_{E}+2\ln \frac{\Lambda }{4\pi T})+O(k^{4}/T^{2}),\\
\Pi_{1}^{\prime}(0)&=&\frac{1}{3(4\pi )^{2}}
(2+2\gamma_{E}+2\ln \frac{\Lambda }{4\pi T}).
\eqa
We also need the self-energy at zero external momentum to 
two loop order.
The contributing diagrams are displayed in Fig.~\ref{selfs}. Many of the
two-loop sum-integral
vanish in dimensional regularization, while others factorize into products
of one-loop sum-integrals. Consider e.g. the first diagram in 
Fig.~\ref{selfs}:
\bqa
4e^{4}\hbox{$\sum$}\!\!\!\!\!\!\int_{PQ}
\frac{p_{0}^{2}(P+K)^{2}}{P^{6}Q^{2}(P-Q)^{2}}.
\eqa
We rewrite this as 
\beq
4e^{4}\hbox{$\sum$}\!\!\!\!\!\!\int_{PQ}\Big[\frac{2p_{0}^{2}}{P^{6}(P-Q)^{2}}
+\frac{2p_{0}^{2}}{P^{4}Q^{2}(P-Q)^{2}}-\frac{p_{0}^{2}}{P^{6}Q^{2}}\Big].
\eeq
Changing variables $Q\rightarrow P-Q$ in the first term, and using the results
for the two-loop sum-integrals in appendix A, we find
\beq
4e^{4}\hbox{$\sum$}\!\!\!\!\!\!\int_{PQ}\frac{p_{0}^{2}}{P^{6}Q^{2}}.
\eeq
The other sum-integrals are reduced similarly and after some 
calculations we find
\beq
\Pi_{2}(0)=-2(d-1)e^{4}\hbox{$\sum$}\!\!\!\!\!\!\int_{PQ}
\frac{1}{P^{4}Q^{2}}
+8(d-1)e^{4}\hbox{$\sum$}\!\!\!\!\!\!\int_{PQ}
\frac{p_{0}^{2}}{P^{6}Q^{2}}.
\eeq
The ultraviolet divergences in the above 
sum-integrals actually cancel, and so we 
are left with a finite expression for $\Pi_{2}(0)$. This cancellation also
takes place in the case of QED~\cite{jens}. Using the tabulated one-loop
sum-integrals in appendix A, one obtains
\beq
\Pi_{2}(0)=\frac{e^{4}T^{2}}{(4\pi)^{2}}.
\eeq
Using these results, we finally obtain $m_{s}^{2}$ to 
order $e^{4}$:
\beq
m_{s}^{2}\approx\frac{e^{2}T^{2}}{3}\Big[1+\Big(\frac{7}{3}-
\frac{2}{3}\gamma_{E}-\frac{2}{3}\ln \frac{\Lambda }{4\pi T}\Big)
\frac{e^{2}}{(4\pi )^{2}}\Big].
\eeq
Note that one could have obtained this result without carrying out
wave function renormalization in Eq.~(\ref{divergent}). Instead one uses 
Eq.~(\ref{scrm}) and the divergence there is canceled by the charge 
renormalization counterterm.

In the effective theory the contributing diagrams are displayed 
in Figs.~\ref{vblob}$-$\ref{sblob} and again the blobs denote 
mass insertions. 
Denoting the self-energy by $\Pi_{E}(k,\Lambda)$, the
screening mass is given by the solution to the equation
\beq
\label{els}
k^{2}+m^{2}(\Lambda )+\Pi_{\scriptsize E}(k,\Lambda)=0,
\hspace{1cm}k^{2}=-m^{2}_{s}.
\eeq
All loop integrals involve
massless fields and these vanish in dimensional
regularization. Hence $\Pi_{E}(k,\Lambda)=0$, and so the matching 
relation becomes 
$m^{2}(\Lambda )\approx m_{s}^{2}$.
Thus 
\bqa
m^{2}(\Lambda)&=&\frac{e^{2}T^{2}}{3}\Big[1+\Big(\frac{7}{3}-
\frac{2}{3}\gamma_{E}-\frac{2}{3}\ln \frac{\Lambda }{4\pi T}\Big)
\frac{e^{2}}{(4\pi )^{2}}\Big].
\eqa 
Using the renormalization group equation for the electric coupling
\beq
\mu\frac{de^{2}}{d\mu}=\frac{e^{4}}{24\pi^{2}}+O(e^{6}),
\eeq
one can verify that the apparent $\Lambda$-dependence of 
$m^{2}(\Lambda)$
is illusory. This implies that, up to correction of order $e^{6}$, one can
trade $\Lambda$ for an arbitrary renormalization scale $\mu$. 
The reason behind this fact is that the physical screening mass does
not receive logarithmic corrections in the effective theory to order $e^{4}$ 
(see section 4).
An important question is how to choose the arbitrary scale
$\Lambda$. The mass parameter has been obtained by integrating out the
nonstatic modes with masses equal to or greater than $2\pi T$, 
so one expects
that $\Lambda =2\pi T$ is an appropriate choice~\cite{braaten}. 
Making this choice, 
the coefficient in front
of $e^{2}/(4\pi )^{2}$ takes the value 2.407.., which is reasonably small.
\section{The Electric Screening Mass}
Now that we have determined the short-distance coefficients we shall 
use the effective three-dimensional
field theory and calculate the electric screening mass.
We shall do so using perturbation theory
and in order to take the physical effect of
screening into account, 
we must now include the mass parameters in the free part of the 
effective Lagrangian.
This corresponds to the 
following partition of ${\cal L}_{\mbox{\scriptsize ESQED}}$:
\bqa \nonumber
({\cal L}_{\mbox{\scriptsize ESQED}})_{0}&=&\frac{1}{4}
F_{ij}F_{ij}+(\partial_{i}
\phi^{\dagger})(\partial_{i}\phi )+M^{2}(\Lambda )\phi^{\dagger}
\phi+\frac{1}{2}(\partial_{i}A_{0})^{2}\\ \nonumber
&&
+\frac{1}{2}
m^{2}(\Lambda )A_{0}^{2}
+{\cal L}_{\mbox{\footnotesize gf}}
+{\cal L}_{\mbox{\footnotesize gh}}, \\
({\cal L}_{\mbox{\scriptsize ESQED}})_{\mbox{\footnotesize int}}&=&
e^{2}_{E}(\Lambda)\phi^{\dagger}\phi (
A_{i}^{2}+A_{0}^{2})-ie_{E}(\Lambda )A_{i}
(\phi^{\dagger}\partial_{i}\phi-\phi\partial_{i}
\phi^{\dagger})+\delta{\cal L}.
\eqa
The physical screening masses are given by the
self-consistent solutions to Eqs.~(\ref{sscr}) and~(\ref{els}).
The solution to 
Eq.~(\ref{sscr}) to
leading order in coupling is equal to the mass parameter 
$M^{2}(\Lambda)$.
However, recently it has been realized [7,19]
that the
this equation has no self-consistent
solution beyond leading order in perturbation theory.
The problem is the last diagram in Fig.~\ref{eskalar}, which 
has a branch point singularity 
at $k=im_{s}$. The problem is the same as in QCD, namely a scalar field
interacting with a massless gauge field in three dimensions.
We shall not discuss this any further, but refer to~\cite{nonp}
where a nonperturbative definition of the scalar screening mass is discussed
in detail.

The one and two-loop diagrams that contribute to the
electric
screening mass in ESQED are displayed in Figs.~\ref{vblob}$-$\ref{sblob},
except that the diagrams with a mass insertion are absent. 
We then find
\bqa\nonumber 
\Pi_{E}\,(k,\Lambda )&
=&2e^{2}_{E}\int_{p}\frac{1}{p^{2}+M^{2}}
-2e^{4}_{E}\int_{pq}\frac{\delta_{ii}}{q^{2}(p^{2}+M^{2})}
\\ \nonumber
&&+2
e^{4}_{E}\int_{pq}\frac{(p+q)^{2}}{({\bf p}
-{\bf q})^{2}(p^{2}+M^{2})^{2}(q^{2}+M^{2})}
-2e^{4}_{E}\int_{pq}\frac{1}{(q^{2}+M^{2})^{2}(p^{2}+m^{2})}\\
&&-4e^{4}_{E}\int_{pq}\frac{1}{(p^{2}+M^{2})
(({\bf p}+{\bf q})^{2}+m^{2})(({\bf p}+{\bf q}+{\bf k})^{2}+M^{2})}.
\eqa
The integrals may be reduced to known ones by algebraic 
manipulations, which
involve change of variables, in analogy with the calculations in the previous
section. The integrals needed are tabulated in appendix B.
The second integral above, which corresponds to the the first graph in
Fig.~\ref{sblob}, vanishes in dimensional regularization due to the 
masslessness of the photon.
Moreover, the last integral is dependent on the external momentum, and
the self-consistent solution to Eq.~(\ref{els}) at 
next-to-leading
order in $e^{2}/m$ is found by evaluating the integral at 
the point $k=im_{s}$.
This can be seen from expanding the integral in powers of $k^{2}$ and noting
that all terms are equally important for $k\sim eT$. 
The calculation of this diagram is carried out in some detail in appendix B.
Notice also that the logarithmic term from this integral is exactly
canceled by a corresponding term in the second two-loop integral above.
Hence, there is no logarithmic correction to the electric screening mass
from the effective theory through order $e^{4}$.
Adding the different pieces, we obtain the physical screening mass 
squared to order
$e^{4}$, which is the main result of the present paper:
\beq
\label{mainres}
m_{s}^{2}=T^{2}\Big[\frac{e^{2}}{3}-\frac{e^{3}}{4\pi}+
\frac{e^{4}}{(2\pi )^{2}}\Big(-1+\frac{1}{2\sqrt{3}}
+(1+\frac{\sqrt{3}}{2})
\ln (1+\frac{2}{\sqrt{3}})\Big)-\frac{2e^{4}}{(12\pi )^{2}}
\Big(1+\gamma_{E}
+\ln\frac{\mu}{4\pi T}\Big)\Big].
\eeq
This result is in accordance with the calculation of 
Blaizot {\it et al.},
who used resummation methods~\cite{parw1}. 
Using the renormalization group equation
for $e$, we find that physical screening mass is independent of the 
renormalization scale $\Lambda$ up to corrections of order $e^{5}$.

We would also like to comment upon the $e^{3}$ term in Eq.~({\ref{mainres}).
It is non-analytic in $e^{2}$, and is the first contribution to the
screening mass squared from momentum scales of order $eT$. 
In resummed perturbation
theory it arises from a one-loop computation which involves the
dressed scalar propagator. An $e^{3}$ term is not present in QED,
since there is no bosonic propagator in the corresponding 
one-loop graph and
that fermions need no resummation because the Matsubara 
frequencies provide
the necessary infrared cut off~\cite{jens}.
\section{Conclusions}
To summarize, we have derived the electric screening mass squared to order 
$e^{4}$ in scalar electrodynamics
from an effective field theory approach. 
This method makes it possible to unravel the contributions to
physical quantities from momentum scales
of order $T$ and $eT$. This greatly simplifies calculations, since there
is only a single scale involved at a time.

The running of the coupling constant first enters at order 
$e^{4}$ for 
the electric screening mass squared. 
Thus, an $e^{4}$ calculation is needed to
determine the appropriate scale $\mu$ in the $e^{2}$ correction to
this quantity. As previously argued, $\mu =2\pi T$ is a 
reasonable and physically motivated choice.
Finally, we would like to outline the calculations necessary to obtain 
the electric screening mass squared to order $e^{5}$.
This requires the matching of the scalar mass parameter 
to two loop order, and 
one also has to find the coefficient for the 
quartic self-interaction of the temporal gauge field. 
In the effective theory
the contribution to the screening mass squared involves
the computation of three-loop diagrams. 
\section*{Acknowledgments}
The author is grateful to F. Ravndal
for beneficial discussions.
\appendix\bigskip\renewcommand{\theequation}{\thesection.\arabic{equation}}
\setcounter{equation}{0}\section{Sum-integrals in the Full Theory}
In this appendix we give the necessary details for
the sum-integrals used in present work. We closely follow
the conventions of Ref.~\cite{braaten}.
Throughout the work we use the imaginary 
time formalism, where the four-momentum is $P=(p_{0},{\bf p})$
with $P^{2}=p_{0}^{2}+{\bf p}^{2}$. 
The Euclidean energy takes on discrete values, $p_{0}=2n\pi T$
for bosons.
Dimensional regularization is used to
regularize both infrared and ultraviolet divergences by working
in $d=4-2\epsilon$ dimensions, 
and we apply the $\overline{\mbox{MS}}$ 
renormalization scheme. 
We use the following shorthand notation
for the sum-integrals that appear below:
\bqa
\hbox{$\sum$}\!\!\!\!\!\!\int_P f(P)&\equiv 
&\Big( \frac{e^{\gamma_{\tiny E}}\mu^{2}}
{4\pi}\Big )^{\epsilon}\,\,\,\,
T\!\!\!\!\!
\sum_{p_{0}=2\pi nT}\int\frac{d^{3-2\epsilon}k}
{(2\pi)^{3-2\epsilon}}f(P).
\eqa
Then $\mu$ coincides with the
renormalization scale in the $\overline{\mbox{MS}}$ renormalization
scheme.

The specific one-loop sum-integrals needed are listed below
\bqa
\hbox{$\sum$}\!\!\!\!\!\!\int_P\frac{1}{P^{2}}&=&\frac{T^{2}}{12}
\Big[1+\Big(2\ln\frac{\mu}{4\pi T}+
2+2\frac{\zeta^{\prime}(-1)}{\zeta (-1)}\Big)\epsilon
+O(\epsilon^{2})\Big],\\
\hbox{$\sum$}\!\!\!\!\!\!\int_P\frac{1}{(P^{2})^{2}}&=&\frac{1}{(4\pi )^{2}}
\Big[\frac{1}{\epsilon}+2\ln\frac{\mu}{4\pi T}
+2\gamma_{E}+O(\epsilon )\Big],\\
\hbox{$\sum$}\!\!\!\!\!\!\int_P\frac{p_{0}^{2}}{(P^{2})^{2}}
&=&-\frac{T^{2}}{24}
\Big[1+\Big(2\ln\frac{\mu}{4\pi T}
+2\frac{\zeta^{\prime}(-1)}{\zeta (-1)}\Big)\epsilon
+O(\epsilon^{2} )\Big],\\
\hbox{$\sum$}\!\!\!\!\!\!\int_P\frac{p_{0}^{2}}{(P^{2})^{3}}&=&
\frac{1}{4(4\pi )^{2}}
\Big[\frac{1}{\epsilon}+2\ln\frac{\mu}{4\pi T}+2+2\gamma_{E}+O(\epsilon)\Big].
\eqa
The easiest way to evaluate the bosonic sum-integrals is 
to perform the integration over $k$, and then do the
frequency sum by employing the integral representation of the
Riemann zeta function~\cite{arnold1}. For instance:
\beq
\hbox{$\sum$}\!\!\!\!\!\!\int_P\frac{1}{(P^{2})^{n}}=T^{d-2n}\frac{2(-1)^{n}
\pi^{\frac{d-1}{2}}}{(2\pi )^{2n}\Gamma (n)}\zeta (2n+1-d)
\Gamma (\frac{2n+1-d}{2}).
\eeq
We also need some two-loop sum-integrals. They can be evaluated by contour
methods~\cite{kapusta}. Arnold and Zhai~\cite{arnold1}, 
and Kastening and Zhai~\cite{kast} have 
calculated and tabulated
all the sum-integrals in the list below, except for the last one.
\bqa
\hbox{$\sum$}\!\!\!\!\!\!\int_{PQ}\frac{1}{P^{2}Q^{2}(P+Q)^{2}}&=&0,\\
\hbox{$\sum$}\!\!\!\!\!\!\int_{PQ}\frac{q_{0}^{2}}{P^{4}Q^{2}(P+Q)^{2}}
&=&\frac{T^{2}}{12(4\pi )^{2}}
+O(\epsilon),\\
\hbox{$\sum$}\!\!\!\!\!\!\int_{PQ}
\frac{p_{0}^{2}}{P^{4}Q^{2}(P+Q)^{2}}&=&0,\\
\hbox{$\sum$}\!\!\!\!\!\!\int_{PQ}
\frac{p_{0}q_{0}}{P^{2}Q^{4}(P+Q)^{2}}&=&0.
\eqa  
The last integral can be found from the others by considering
\beq
\hbox{$\sum$}\!\!\!\!\!\!\int_{PQ}
\frac{(p_{0}+q_{0})^{2}}{P^{4}Q^{2}(P+Q)^{2}},
\eeq
and then changing variables, $Q^{\prime}=P+Q$. 
\setcounter{equation}{0}\section{Integrals in the Effective Theory}
In the effective three-dimensional theory we use dimensional regularization
in $3-2\epsilon$ dimensions to regularize infrared and ultraviolet 
divergences.
In analogy with Appendix A, we define
\beq
\int_{p}f(p)\equiv\Big( \frac{e^{\gamma_{\tiny E}}\mu^{2}}
{4\pi}\Big )^{\epsilon}\int\frac{d^{3-2\epsilon}p}
{(2\pi)^{3-2\epsilon}}f(p).
\eeq
Again $\mu$ coincides with the renormalization scale in the 
modified minimal subtraction renormalization
scheme.

In the effective theory we need the following one-loop integrals
\bqa
\int_{p} \frac{1}{p^{2}+m^{2}}&=&-\frac{m}{4\pi}
\Big[1+\Big(2\ln\frac{\mu}{2m}+2\Big)\epsilon
+O(\epsilon^{2})\Big],\\
\int_{p} \frac{1}{(p^{2}+m^{2})^{2}}&=&\frac{1}{8\pi m}
\Big[1+\Big(2\ln\frac{\mu}{2m}\Big)\epsilon
+O(\epsilon^{2})\Big].
\eqa
All integrals are
straightforward to evaluate in 
dimensional regularization. Details may be found in Ref.~\cite{ryder}.
The specific two-loop integrals needed are
\bqa\nonumber
\int_{pq} \frac{1}{(p^{2}+m^{2})(q^{2}+m^{2})({\bf p}-{\bf q})^{2}}&=&
\frac{1}{(4\pi )^{2} m}
\Big[\frac{1}{4\epsilon}
+\frac{1}{2}\\&&
+\ln\frac{\mu}{2m}+O(\epsilon )\Big],\\
\int_{pq} \frac{1}{(p^{2}+m^{2})^{2}(q^{2}+m^{2})({\bf p}-{\bf q})^{2}}&=&
\frac{1}{(4\pi )^{2} m^{2}}
\Big[\frac{1}{4}+
O(\epsilon )\Big],\\  \nonumber
\label{dimjens}
\left.\int_{pq} \frac{1}{(p^{2}+M^{2})(q^{2}
+M^{2})[({\bf p}+{\bf q}+{\bf k})^{2}+m^{2}]}\right|_{k=im}
&=&
\frac{1}{(8\pi )^{2}}\Big[\frac{1}{\epsilon}
+6\\
-
4\ln [\frac{2(M+m)}{\mu}]
+4\frac{M}{m}\ln\frac{M}{M+m}+O(\epsilon )\Big].&&
\eqa
The two first of these integrals can be found in Ref.~\cite{braaten3}.
The integral in Eq.~(\ref{dimjens}) has previously been calculated by
Braaten and Nieto for $m=M$ in Ref.~\cite{braaten}. Below we compute it
for the more general case $m\neq M$ using their methods.
The integral can best be computed by going to coordinate space. The 
Fourier
transform of the propagator is
\beq
V_{m}(R)=\int_p e^{i{\bf p}\cdot {\bf R}}\,\frac{1}{p^{2}+m^{2}}.
\eeq
It can be expressed in terms of a modified Bessel function
\beq
V_{m}(R)=\Big(\frac{e^{\gamma_{E}}\mu^{2}}{4\pi}\Big)
^{\epsilon}\frac{1}{(2\pi )^{3/2-\epsilon}}
\Big(\frac{m}{R}\Big)^{1/2-\epsilon}K_{1/2-\epsilon}(mR).
\eeq
In three dimensions ($\epsilon =0$) this is the Yukawa potential:
\beq
\tilde{V}_{m}(R)=\frac{e^{-mR}}{4\pi R}.
\eeq
For small $R$ it can be written as a sum of two Laurent series in $R^{2}$.
One of these is singular beginning with an $R^{-1+2\epsilon}$ 
term and the other
is regular which begins with an $R^{0}$ term:
\bqa
V_{m}(R)&=&\Big(\frac{e^{\gamma_{E}}\mu^{2}}{4}\Big)
^{\epsilon}\frac{\Gamma (\frac{1}{2}-\epsilon)}
{\Gamma (\frac{1}{2})}\frac{1}{4\pi}R^{-1+2\epsilon}
\Big[1+\frac{m^{2}R^{2}}{2(1+2\epsilon)}+O(m^{4}R^{4})\Big]\\
&&-(e^{\gamma_{E}}\mu^{2})^{\epsilon}
\frac{\Gamma (-\frac{1}{2}+\epsilon)}
{\Gamma (-\frac{1}{2})}\frac{1}{4\pi}m^{1-2\epsilon}
\Big[1+\frac{m^{2}R^{2}}{2(3-2\epsilon)}+O(m^{4}R^{4})\Big].
\eqa
The integral can be written
\beq
\int_{pq} \frac{1}{(p^{2}+M^{2})(q^{2}+M^{2})
[({\bf p}+{\bf q}+{\bf k})^{2}+m^{2}]}
=\int_{R} e^{i{\bf kR}}V^{2}_{M}(R)V_{m}(R)
\eeq
The radial integration is now split into two regions, 
$0<R<r$ and $r<R<\infty$.
The ultraviolet divergences arise from the region $R\rightarrow 0$. 
This implies that we can set $\epsilon =0$ in the region 
where $r<R<\infty$.
Hence, one can write the integral as
\bqa
\int e^{i{\bf kR}}V^{2}_{M}(R)V_{m}(R)&=&
\Big(\frac{e^{\gamma_{E}}\mu^{2}}{2k}\Big)^
{-\epsilon}\frac{(2\pi)^{3/2}}{\sqrt{k}}
\int_{0}^{r}dRR^{3/2-\epsilon}J_{1/2-\epsilon}(kR)V^{2}_{M}(R)
V_{m}(R)\\
&&+\frac{4\pi}{k}\int_{r}^{\infty}dRR\sin (kR)\tilde{V}^{2}_{M}(R)
\tilde{V}_{m}(R).
\eqa
Here, $J_{\nu}(x)$ is an ordinary Bessel function. The Bessel function has 
the following expansion for small $R$:
\beq
J_{1/2-\epsilon}(kR)=\frac{1}{\Gamma (\frac{3}{2}-\epsilon)}
\Big(\frac{1}{2}kR\Big)^{1/2-\epsilon}[1+O(k^{2}R^{2})].
\eeq
Using this expansion and the small $R$ expansion of the 
potential, the first
integral is, after dropping terms that vanish in the limit 
$r\rightarrow 0$
\bqa \nonumber
\Big(\frac{e^{\gamma_{E}}\mu^{2}}{2k}\Big)
^{-\epsilon}\frac{(2\pi)^{3/2}}{\sqrt{k}}
\int_{0}^{r}dRR^{3/2-\epsilon}J_{1/2-\epsilon}
(kR)V^{2}_{M}(R)V_{m}(R)&=&
\frac{1}{(8\pi )^{2}}\Big[\frac{1}{\epsilon}+
4\ln \mu r\\
&&+2+4\gamma_{E}\Big]+O(\epsilon ).
\eqa
The second integral can be found in e.g Ref.~\cite{grad} and equals
\beq
\frac{i}{2k(4\pi)^{2}}\Big[(2M+m+ik)\Gamma [-1,(2M+m+ik)r]
-(2M+m-ik)\Gamma [(2M+m-ik)r]\Big].
\eeq
Evaluating this at $k=im$ yields
\bqa \nonumber
\left.\frac{4\pi}{k}\int_{r}^{\infty}dRR\sin (kR)\tilde{V}^{2}_{M}(R)
\tilde{V}_{m}(R)\right|_{k=im}&=&
\frac{1}{(4\pi )^{2}}\Big[\frac{M}{m}\ln\frac{M}{M+m}-\gamma_{E}+1
\\
&&-\ln [2(M+m)r]\Big]+O(\epsilon),
\eqa
where we have used the series expansion of the incomplete gamma 
function
\beq
\Gamma [-1,x]=\frac{1}{x}+\gamma_{E}-1+\ln x +O(x^{2}),
\eeq
and dropped terms that vanish as $r\rightarrow 0$.
Collecting our results we obtain Eq.~(\ref{dimjens}).
The logarithms of $r$ cancel and our result
reduces to the one found in Ref.~\cite{braaten} 
in the case $m=M$, as it should.
 
\pagebreak
\begin{figure}[b]
\underline{FIGURE CAPTIONS:}
\caption{One-loop self-energy diagrams for the scalar field in SQED.}
\label{1skalar}
\caption{One-loop self-energy corrections for $\phi$ in ESQED.}
\label{eskalar}
\caption{One-loop self energy diagrams for the timelike 
photon in SQED.}
\label{selfa}
\caption{Two-loop self energy diagrams for the timelike 
photon in SQED.}
\label{selfs}
\caption{One-loop self-energy corrections for $A_{0}$ in ESQED.}
\label{vblob}
\caption{Two-loop self-energy corrections for $A_{0}$ in ESQED.}
\label{sblob}
\end{figure}
\pagebreak
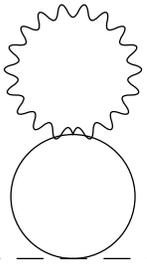
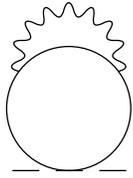
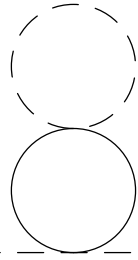
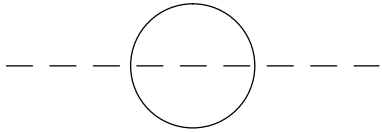
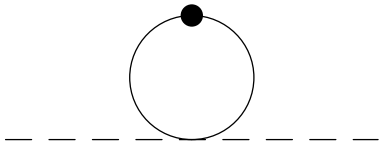
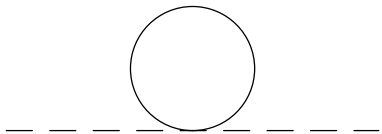
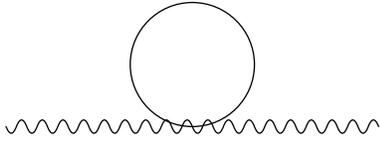
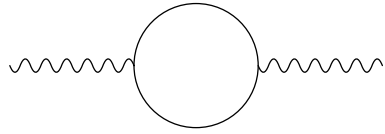
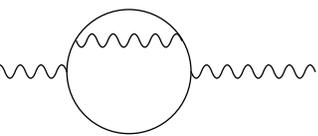
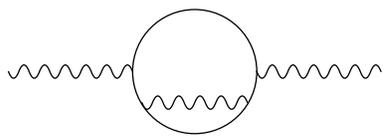
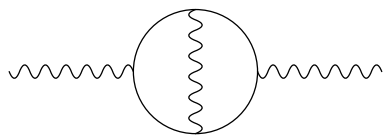
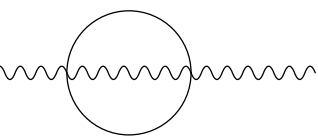
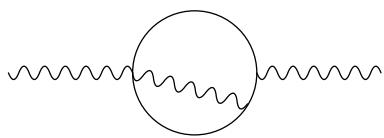
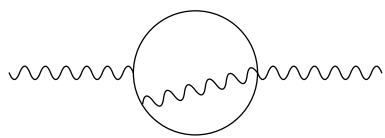
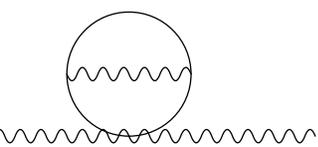
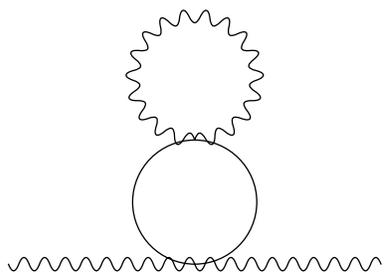
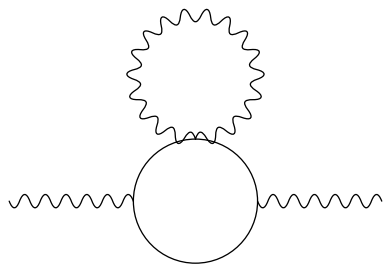
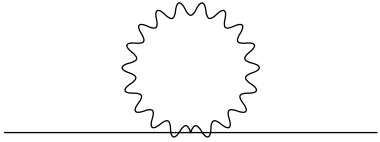
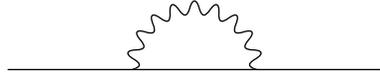
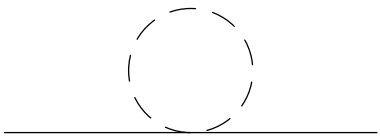
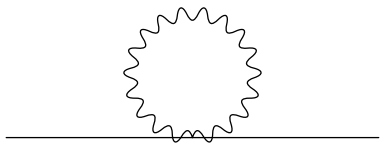
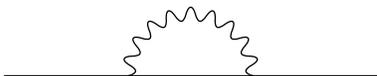
\end{document}